\documentclass{ar-1col-S2O}
\usepackage[numbers]{natbib}
\usepackage{url}
\jname{Annu. Rev. Condens. Matter Phys.}
\jvol{AA}
\jyear{YYYY}
\doi{10.1146/((please add article doi))}

\usepackage{mathtools}
\usepackage{bbm}
\usepackage{ulem}
\usepackage{braket}
\usepackage{bbold}
\usepackage{comment}
\usepackage{amsthm} % Theorem Formatting
\usepackage{amssymb}    % Math symbols such as \mathbb
\usepackage{bm}
\usepackage{multirow}
\usepackage{makecell}
\usepackage{slashed}

\renewcommand{\sc}[1]{\mathcal{#1}}

\def\so{\mathrm{SO}}
\def\mod{\text{ mod }}

\def\U{\mathrm{U}(1)}

\def\bZ{\mathbb{Z}}

\newtheorem{theorem}{Theorem}[section]

\theoremstyle{definition}

\usepackage[colorlinks, linkcolor=blue, citecolor=blue, urlcolor=blue]{hyperref}

\begin{document}

% Page header
\markboth{Zou \& Cheng}{LSM Anomalies and Anomaly Matching}

% Title
\title{Lieb-Schultz-Mattis Anomalies and Anomaly Matching}

% Authors, affiliations
\author{
Liujun Zou$^1$ and Meng Cheng$^2$
\affil{$^1$Department of Physics, National University of Singapore, 117551, Singapore; email: lzou@nus.edu.sg}
\affil{$^2$Department of Physics, Yale University, New Haven, Connecticut 06511, USA; email: m.cheng@yale.edu
}
}

% Abstract
\begin{abstract}
Lieb-Schultz-Mattis (LSM) anomalies are powerful symmetry-based constraints on the correlation, entanglement and dynamics of quantum many-body systems. In this review, we discuss various LSM anomalies and anomaly matching. We start with a pedagogical introduction to these subjects in quantum spin chains, and then generalize the discussion to higher dimensions and other systems. Besides covering the topics related to the standard LSM anomalies, we also review LSM anomalies in disordered systems where the lattice symmetries are only preserved on average, fermionic systems, and systems where the symmetric short-range entangled states are possible but must be nontrivial symmetry-protected topological phases.
\end{abstract}

% Keywords
\maketitle

% Table of Contents
\tableofcontents

\section{INTRODUCTION}

Understanding complex quantum many-body systems is an important but notoriously challenging task, as most models of such systems cannot be analytically solved and numerical approaches are limited in various ways. Therefore, Lieb-Schultz-Mattis-type (LSM) constraints become highly valuable \cite{LSM}, which allow one to get important information about a system without solving its Hamiltonian. The typical statement of an LSM constraint starts with certain symmetry conditions where the symmetries include some lattice symmetries, and concludes that all states satisfying these conditions must have some nontrivial patterns of correlation and entanglement, and all local and even some non-local Hamiltonians obeying these conditions cannot have a unique gapped ground state (under periodic boundary conditions). Such LSM constraints have wide applicability, and the difference between different versions of them lies in the dimensionality of the system, the type of the symmetries, whether the system is a spin system or fermionic system, etc. An incomplete list of LSM constraints, some of which will be reviewed below, can be found in Refs. \cite{LSM, OshikawaLSM, HastingsLSM, PoPRL2017, Watanabe2018, Cheng2018, Ogata2018, Else_LSM_2019, Yao:2020xcm, Ogata2020, Ma2022, Kapustin2024, Liu2024a, Ma2024, Liu2025, LuRanOshikawa2020, Lu2024, WuHoLu2017, XuSPTLSM, JiangSPTLSM, Kobayashi2018, Pace2024, Pace2025, Pace2026, Anakru2026, Ning2026}.

Recently, LSM constraints have been interpreted as 't Hooft anomalies \cite{ChengPRX2016, JianPRB2018, ChoPRB2017, MetlitskiPRB2018, Ye2021, Cheng2022, Seifnashri2023, Liu2024}, so we will use ``LSM constraints" and ``LSM anomalies" interchangeably. The notion of 't Hooft anomalies originates from quantum field theory (QFT), where one considers a relativistic QFT with a certain internal symmetry $G$ \cite{tHooft:1979rat}. Having the symmetry means that the theory can be coupled to a classical, background $G$ gauge field. The anomaly arises when the theory with the background gauge field cannot be made gauge-invariant\footnote{More precisely, the gauge non-invariance cannot be remedied by adding local counterterms.}. Consequently, the gauge field cannot be made dynamical, although the original theory is a consistent theory itself. Thus 't Hooft anomalies in QFTs can be defined as an obstruction to gauging a symmetry. A well-known example of 't Hooft anomalies is the chiral anomaly of the axial/chiral U(1) symmetry of a Dirac fermion in odd spatial dimensions  \cite{Adler1969, Bell1969}.

In the past decades, the study of symmetry-protected topological (SPT) phases in condensed matter physics \cite{Chen:2011pg} has led to significant advances in the understanding of 't Hooft anomalies, beyond their original formulation in relativistic QFTs. In particular, it has been understood that quantum anomalies can be directly defined in lattice systems using the symmetry actions, where the symmetries may be internal or involve spatial transformations \cite{CZX, ElsePRB2014, Kapustin2024, Kawagoe2025, Liu2025}. Compared to the continuum version, the lattice setting poses additional challenges in understanding these anomalies. For example, it is unclear how to make sense of background gauge fields for spatial symmetries. A completely general definition, characterization, and classification of ``lattice anomalies" is still an active research area, and we will not pursue this direction here (see Refs. \cite{Kapustin:2025nju, LatticeAnomaly1, Shirley:2025yji, Czajka2025} for recent developments).

A key fact shared by all instances of 't Hooft anomalies is that
the anomaly is a \emph{``kinematic"} property, independent of the details of energetics determined by the Hamiltonian. In fact, as we will see, the anomaly is entirely determined by the structure of the underlying local Hilbert spaces and how the symmetries act. This is why a nontrivial anomaly puts strong constraints on low-energy dynamics and forbids the existence of a symmetry-preserving, short-range entangled (SRE) state in the system. This observation leads to the powerful \emph{``anomaly matching"} framework to analyze which low-energy theories can emerge in a microscopic system with a given symmetry condition. 

The goal of this review is to give an elementary introduction to LSM anomalies and anomaly matching. In Sec. \ref{sec: 1D}, we discuss in detail how the LSM constraints manifest themselves in quantum spin chains and demonstrate the anomaly matching. In Sec. \ref{sec: higher dimensions}, we discuss how the LSM constraints discussed in Sec. \ref{sec: 1D} extend to higher dimensions. In Sec. \ref{sec: anomaly matching}, we describe the systematic framework for anomaly matching, and sketch how it is applied to different example systems. In Sec. \ref{sec: other generalizations}, we discuss various other generalizations of LSM constraints. In Sec. \ref{sec: crystalline SPTs}, we discuss crystalline SPTs that are useful for understanding the LSM anomalies.

\section{LSM ANOMALY IN QUANTUM SPIN CHAINS} \label{sec: 1D}

In this section, we discuss the LSM anomaly in quantum spin chains, where LSM constraints were originally discovered \cite{LSM}. We use it as a prototypical example to demonstrate different perspectives on the LSM theorem as a manifestation of an anomaly, its various generalizations, the rigorous formulation, and anomaly matching between the UV system and IR theory.

\subsection{The LSM Theorem for Spin Chains} \label{subsec: LSM for spin-1/2 chain}

We consider, as an example, the antiferromagnetic spin-$S$ Heisenberg chain
\begin{equation}
    H=\sum_{j=1}^L \bm{S}_j\cdot \bm{S}_{j+1} + \cdots,
    \label{eq:Heisenberg-chain}
\end{equation}
with periodic boundary conditions $\bm{S}_{j+L}=\bm{S}_j$. The dots include other interactions invariant under SO(3) spin rotations and lattice translations. A version of the LSM theorem is the following statement \cite{LSM}, which applies to any local Hamiltonian of the form in Eq. \eqref{eq:Heisenberg-chain}.

\begin{theorem} \label{thm: original LSM}
The Hamiltonian in Eq. \eqref{eq:Heisenberg-chain} cannot have both a gapped spectrum and a unique ground state, if $S\in\mathbb{Z}+\frac{1}{2}$.
\end{theorem}

The proof of the theorem proceeds as follows: Suppose the ground state is unique (otherwise the statement already holds). Then it must be a spin-singlet. Next we construct an orthogonal state whose energy is bounded by $O(L^{-1})$, which shows that the system is gapless and completes the proof. To do this, define the following ``twist operator"
\begin{equation}
    U_{2\pi}=\exp\Big(\frac{2\pi i}{L}\sum_{j=1}^L jS_j^z\Big).
\end{equation}
Now consider the state $\ket{\psi_1}=U_{2\pi}\ket{\psi_0}$. It is straightforward to show that $\braket{\psi_1|H|\psi_1}-\braket{\psi_0|H|\psi_0}=O(L^{-1})$ (the details can be found in standard texts, e.g., Ref. \cite{auerbach1994interacting}).

The key step then is to prove that $\ket{\psi_1}$ is \emph{orthogonal} to $\ket{\psi_0}$.  It is in this step that we need to use the value of spin per site. Specifically, one can show that the lattice momentum of $|\psi_1\rangle$ differs from that of $|\psi_0\rangle$ by $\pi$ when $S\in\mathbb{Z}+\frac{1}{2}$, which means $|\psi_1\rangle$ and $|\psi_0\rangle$ are orthogonal as they are eigenstates of the translation operator with different eigenvalues.

To further understand the physical meaning of the twist operator and to connect with modern perspectives on 't Hooft anomalies, we follow Ref. \cite{OshikawaLSM} and study the following family of Hamiltonians:
\begin{equation}
    H'[\alpha]=\sum_{j=1}^L (e^{\frac{i\alpha}{L}}S_j^+ S_{j+1}^- + \text{h.c.}) + S_j^zS_{j+1}^z.
    \label{eq:Ham-theta-1}
\end{equation}
We may think of $H'[\alpha]$ as coupling the spin chain to a background gauge field of the $\U$ symmetry of $S_z$ conservation (a subgroup of SO(3)).  In particular, $H'[\alpha=2\pi]$ is unitarily equivalent to $H$:
\begin{equation}
    U_{2\pi}HU_{2\pi}^{-1}=H'[2\pi].
\end{equation}
In other words, when $\alpha=2\pi$ the gauge field is actually trivial, i.e., it is a pure gauge. The gauge transformation is precisely implemented by $U_{2\pi}$.

In this presentation, $H'[\alpha]$ commutes with the lattice translation. A gauge-equivalent presentation is
\begin{equation}
\label{Htheta}
    H[\alpha]=\sum_{j=1}^{L-1} (S_j^+ S_{j+1}^- + \text{h.c.})+(e^{-i\alpha}S_L^+S_1^- + \text{h.c.})+ \sum_{j=1}^LS_j^zS_{j+1}^z,
\end{equation}
where the phase factor is concentrated on one link. Here $H$ is manifestly periodic in $\alpha$: $H[\alpha+2\pi]=H[\alpha]$, while $H'$ is not: $H'[\alpha+2\pi]=U_{2\pi}H'[\alpha]U_{2\pi}^{-1}$. On the other hand, $H$ is no longer invariant under the ``bare" translation, but one can define a modified translation operator $T(\alpha)=e^{i\alpha S_1^z}T$, which commutes with $H[\alpha]$. It satisfies
\begin{equation} \label{eq: modified translation}
    T(\alpha)^L = e^{i\alpha S^z},\quad T(\alpha+2\pi)=-T(\alpha).
\end{equation}

We now give an intuitive explanation of the LSM theorem. Suppose that for $\alpha=0$, the Hamiltonian is gapped with a unique ground state. Then adiabatically turning on $\alpha$ from $0$ to $2\pi$, the defect Hamiltonian $H[\alpha]$ still has a unique gapped ground state \cite{WatanabePRB2018}, denoted by  $\ket{\psi_0(\alpha)}$. The ground state eigenvalue $T_0(\alpha)=\braket{\psi_0(\alpha)|T(\alpha)|\psi_0(\alpha)}$ must be a continuous, single-valued function of $\alpha$. Furthermore, since $S^z=0$ in the ground state, the first relation in Eq. \eqref{eq: modified translation} implies $T_0(\alpha)^L=1$, hence $T_0(\alpha)$ takes quantized values independent of $\alpha$. This is in contradiction with the second relation of Eq. \eqref{eq: modified translation}, which requires $T_0(\alpha)$ to change. The only way to satisfy both is to have level crossing(s) between the ground state and excited states for some values of $\alpha$. This contradiction again confirms the LSM theorem.

We can further show that the level crossing must happen at $\alpha=\pi$. We define $Z=e^{i\pi S^z}, X=e^{i\pi S^x}$. The Hamiltonian $H[\alpha=\pi]$ has a $Z$ defect, with the modified translation $T_Z=e^{i\pi S_1^z}T$. It immediately follows that $T_Z X=-X T_Z$, and hence the defect ground state at $\alpha=\pi$ must be degenerate, i.e., there is level crossing between the ground state and excited states. 

Closely inspecting the above arguments, we have only used the ${\rm O}(2)=\U\rtimes \bZ_2$ subgroup of SO(3). The same is true in the original proof. In fact, the above argument applies even if the symmetry is further reduced to $\bZ_2\times \bZ_2$ (the group of $\pi$ rotations), because we have just proven that the ground state with a $Z$ defect must be degenerate, which contradicts the expectation that when the Hamiltonian is fully gapped with a non-degenerate ground state, the defect Hamiltonian should also be so. In fact, not only the ground state, every energy level of $H[\alpha=\pi]$ must be degenerate, which is a manifestation that the anomaly is a kinematic property.

We can now state a general version of the LSM theorem in 1D with an internal symmetry and the lattice translation symmetry. Suppose the internal symmetry group of the system is $G$. $G$ is ``on-site", which means that the unitary transformation $U(g)$ for $g\in G$ takes the following form: $U(g)=\prod_{j}U_j(g)$, where $U_j(g)$ acts on the $j$-th site and identity elsewhere. We assume that all the symmetry actions are translation invariant, that is, all sites transform in the same way.

By definition, $\{U_j(g)\}_{g\in G}$ forms a representation of $G$. In quantum mechanics, the representation may be projective, that is, $U_j(g)U_j(h)$ is equal to $U_j(gh)$ up to a phase:
\begin{equation}
    U_j(g)U_j(h)=\omega(g,h)U_j(gh).
\end{equation}
A representation is projective when the phase factors $\omega(g,h)$ cannot be ``gauged away", meaning that it cannot be written as $u(g)u(h)/u(gh)$ for some phase factors $u$ (otherwise one can redefine $U_j(g)\rightarrow u(g)U_j(g)$ to eliminate $\omega$). A familiar example is the half-integer-spin representations of SO(3).

We now present the general LSM theorem in this setup:
\begin{theorem} \label{thm: LSM more general}
  A translation-invariant, $G$-symmetric local Hamiltonian with a nontrivial projective representation of $G$ per unit cell cannot have a non-degenerate, gapped ground state.
\end{theorem}

After ruling out a unique gapped ground state by the LSM theorem, one may wonder whether one can further constrain the ground state properties. For spin chains with $\so(3)$ and translation symmetries, on general grounds, one expects two kinds of low-energy dynamics: either the spin excitations are gapless, or the translation symmetry is spontaneously broken (as in, e.g., a dimerized phase). The following theorem from Ref. \cite{Aizenman2001BoundedFluctuations} formalizes this intuition.

\begin{theorem}
\label{thm: dichotomy}
The setup is the same as in Theorem \ref{thm: original LSM}.
    Define $F(a,b)=\langle\big(\sum_{j\in [a,b]}S_j^z)^2\rangle$ as the variance of the total $S^z$ in an interval $[a,b]$. In the thermodynamic limit, if $F(a,b)$ is bounded as $|b-a|$ increases, the ground state must spontaneously break the lattice translation symmetry.
\end{theorem}

Roughly speaking, the condition that $F(a,b)$ is bounded means that the spin correlation function $\braket{S^z_i S^z_j}$ cannot decay too rapidly; if $\braket{S^z_i S^z_j}\sim |i-j|^{-\alpha}$, then one requires $\alpha\leqslant 2$ in order for translation symmetry to not be spontaneously broken.
The well-known example of the nearest-neighbor Heisenberg chain saturates the above bound: Since $\braket{S^z_iS^z_j}\sim |i-j|^{-2}$, the bipartite fluctuation grows logarithmically: $F(a,b)\sim \ln |b-a|$, which is unbounded. Indeed, translation is not spontaneously broken in this case. To show that the translation symmetry requires the spin-spin correlation to not decay too fast, alternatively, one can directly work with the spin correlation functions to prove a similar result \cite{Kimchi2017}.

We remark that translation is not the only lattice symmetry that can lead to a mixed anomaly with internal symmetries. The space group of a 1D chain is quite simple. It is generated by translation and reflection. Reflection can also lead to an LSM anomaly. To see it, we first consider a site-centered reflection ${\cal R}_s$ and discuss an example of the LSM theorem that only requires SO(3) and ${\cal R}_s$. For simplicity, still consider the spin chain Hamiltonian on a periodic chain of $L$ sites. The reflection transformation ${\cal R}_{s}$ centered at site $1$ maps $j$ to $2-j$ mod $L$. We now argue that as long as the Hamiltonian preserves both SO(3) and ${\cal R}_s$, the ground state cannot be non-degenerate.

To see why this is the case, we generalize the argument above, introducing a $Z$ defect on the bond $(1,2)$ (modifying the Hamiltonian to $H[\alpha=\pi]$ in Eq. \eqref{Htheta}). Naively, the presence of the defect breaks ${\cal R}_s$, but ${\cal R}_s'=e^{i\pi S_1^z}{\cal R}$ is a symmetry. However, $X{\cal R}_s'=-{\cal R}_s'X$, so the defect Hamiltonian has degenerate ground states protected by $\mathcal{R}_s'$ and spin rotations, which is incompatible with a non-degenerate ground state.

As before, the above physical argument relies on the assumption that the defect Hamiltonian should remain non-degenerate, if the starting Hamiltonian is so. A rigorous proof of the above LSM constraint was given in Ref. \cite{Ogata2020}.

On the other hand, in a spin-1/2 lattice with a bond-centered reflection symmetry ${\cal R}_b: j\rightarrow 1-j$ mod $L$, one can easily construct a spin-singlet, $\mathcal{R}_b$-invariant SRE state by pairing the spins on site $2j$ and $2j+1$ into a singlet. So there is no LSM anomaly associated with bond-centered reflection and spin rotation. As a lesson, we see that the position of the degrees of freedom (DOF) can determine whether there is an LSM constraint. How should one determine and characterize the LSM constraints in different lattice systems? An answer to this question is provided by lattice homotopy \cite{PoPRL2017, Else_LSM_2019}, which will be reviewed in Sec. \ref{sec: higher dimensions}.

\subsection{Rigorous Approaches}
\label{sec:rigorous}

The original proof for the antiferromagnetic Heisenberg chain in Ref. \cite{LSM} is fully rigorous. However, the physical argument presented in Sec. \ref{subsec: LSM for spin-1/2 chain} for discrete symmetry groups relies on the unproven assumption that the bulk gap remains robust upon introducing a symmetry defect (this has been rigorously established for a U(1) symmetry defect \cite{WatanabePRB2018}).

A rigorous, general proof can be obtained using the powerful theory of operator algebra. Using this approach, Refs. \cite{Liu2024a, Liu2025} proved the following theorem that concerns the interplay between correlation, entanglement and symmetries.

\begin{theorem} \label{thm: general LSM for states}
In a spin chain with symmetry group $G_{\rm int}\times\bZ$, where $G_{\rm int}$ is an on-site unitary or anti-unitary symmetry and $\bZ$ represents the translation symmetry, if the DOF at each site transforms in a projective representation under $G_{\rm int}$, then any symmetric state must either have long-range correlation or violate the entanglement area law.
\end{theorem}

Here having long-range correlation means that the connected correlator does not decay to zero for two local operators that are far away, and the entanglement area law means that the von Neumann entropy for a segment of the spin chain is upper bounded when the length of the segment increases. This theorem means that if a system has an LSM anomaly, then any symmetric state must have some nontrivial correlation and entanglement. In particular, it implies that any symmetric state must be long-range entangled (LRE). We remark that the states discussed here do not have to be the ground state of any natural Hamiltonian. Similar theorems were also proved in Refs. \cite{Ogata2018, Ogata2020}.

What is the relation between Theorems \ref{thm: original LSM}, \ref{thm: LSM more general} and Theorem \ref{thm: general LSM for states}? In fact, Theorems \ref{thm: original LSM} and \ref{thm: LSM more general} can be derived and generalized using Theorem \ref{thm: general LSM for states} and various other theorems. It is known that if a gapped local Hamiltonian has a unique ground state, then this ground state does not have long-range correlation \cite{Hastings2005}. Moreover, this ground state satisfies the entanglement area law \cite{Hastings2007}. So under the symmetry condition in Theorem \ref{thm: general LSM for states}, a symmetric gapped local Hamiltonian cannot have a unique ground state, which proves Theorems \ref{thm: original LSM} and \ref{thm: LSM more general}.

Interestingly, the locality condition in Theorems \ref{thm: original LSM} and \ref{thm: LSM more general} can be relaxed. As proved in Refs. \cite{Liu2024a, Kuwahara2019}, suppose a Hamiltonian is non-local but has interactions decaying faster than $1/r^2$, where $r$ is the distance between the interacting DOF. If it is gapped and has a unique ground state, then its ground state satisfies the entanglement area law. Furthermore, Ref. \cite{Hastings2005} shows that such a ground state does not have long-range correlation. Therefore, we reach the following generalization of Theorems \ref{thm: original LSM} and \ref{thm: LSM more general}.

\begin{theorem} \label{thm: general LSM for spectrum}
Under the same symmetry condition as in Theorem \ref{thm: general LSM for states}, if a Hamiltonian is symmetric and its interactions decay faster than $1/r^2$, with $r$ the distance between the interacting DOF, then this Hamiltonian cannot have a unique gapped ground state.
\end{theorem}

We remark that the above way of thinking about the LSM theorem highlights the LSM anomaly as a constraint on the patterns of correlation and entanglement of the symmetric states, and the constraint on the spectra of symmetric Hamiltonians is its derived consequence. This is a powerful perspective, because Theorem \ref{thm: general LSM for states} only takes the symmetry condition as the input and makes no reference to any Hamiltonian, which underscores the kinematic nature of anomalies. Also, this approach leads to constraints on the spectra of {\it non-local} Hamiltonians under a wide class of symmetries, as stated in Theorem \ref{thm: general LSM for spectrum}, which has not been obtained using other approaches (for the special case with $G_{\rm int}={\rm SO(3)}$, Theorem \ref{thm: general LSM for spectrum} was also proved in Ref. \cite{Ma2024} by explicitly constructing low-lying states).

\subsection{LSM Anomaly Matching in Low-Energy Theory}

\label{subsec: anomaly matching 1d}

An LSM anomaly, by definition, rules out a completely trivial ground state. Its power, however, goes further than just the no-go theorem: The anomaly in fact puts strong constraints on the low-energy theory, known as ``anomaly matching". In this section, we demonstrate anomaly matching using the IR field theory of the spin-1/2 chain as an example.

It is well-known that the IR theory of the XXZ spin chain is a Luttinger liquid \cite{Sachdev_2011}, given by the following Lagrangian (in Euclidean signature):
\begin{align}
	\mathcal{L}&=\frac{i}{2\pi}\partial_\tau\phi \partial_x\vartheta +  {\cal H} , \\
    {\cal H}&=\frac{v}{4\pi}[K^{-1}(\partial_x\vartheta)^2+K(\partial_x\phi)^2]\\
    \vartheta &\sim \vartheta +2\pi , \qquad \phi \sim \phi +2\pi.
    \label{eq: LL}
\end{align}
Here ${\cal H}$ is the Hamiltonian density and $v$ is the velocity. Physically, one can identify $e^{i\vartheta}\sim (-1)^jS_j^+$ as the XY order parameter, $\cos\phi\sim(-1)^jS^z_j$ as the staggered field and $\sin\phi\sim(-1)^j\vec S_j\cdot\vec S_{j+1}$ as the dimer order parameter \cite{Sachdev_2011}.
$R=\sqrt{K}$ is often referred to as the boson radius in field theory literature and is determined by the ZZ anisotropy.   In particular, the Heisenberg model corresponds to $R=1$, in which case the theory is also equivalent to the SU(2)$_1$ WZW conformal field theory (CFT).

For a generic value of $R$, the global internal symmetry of $\mathcal{L}$ is
\begin{equation}\label{Gconee}
G_{c=1}=\big(\U_m\times \U_w\big)\rtimes \bZ_2^C.
\end{equation}
The two U(1) factors are generated by charges $Q_m$ and $Q_w$, whose eigenvalues are quantized to integers.  Following the CFT terminology, the labels stand for momentum and winding, respectively. Suppose that the system has length $L$ and periodic boundary conditions. Then
\begin{equation}
    Q_m = \frac{1}{2\pi}\int_0^L dx\, \partial_x\phi,\quad Q_w = \frac{1}{2\pi}\int_0^L dx\, \partial_x\vartheta.
\end{equation}
The $\bZ_2^C$ factor is generated by the charge conjugation, under which we have $\phi\rightarrow -\phi, \vartheta\rightarrow -\vartheta$ and $(Q_m, Q_w)\rightarrow (-Q_m, -Q_w)$.

Since the field theory arises as the deep IR description of the spin chain, it is important to understand how symmetries in the spin chain are implemented in the field theory. First, one can show that the O(2) spin rotation in the lattice model becomes $\U_m\rtimes \bZ_2^C$ in the field theory, where the $S^z$ density corresponds to $\frac{1}{2\pi}\partial_x\phi$.

More interestingly, one finds that the lattice translation becomes $e^{i\pi Q_m}e^{i\pi Q_w}e^{iPa}$. Here $P$ is the momentum operator of the field theory and $a$ is the lattice spacing, so $e^{iPa}$ implements a continuous translation by $a$. Consequently, $e^{i\pi Q_m}e^{i\pi Q_w}$ is an exact symmetry in the low-energy theory, unlike the other elements of $\U_w$. More precisely, $\U_w$ is only emergent in the low-energy theory and is violated by irrelevant perturbations, except for its $\bZ_2$ subgroup. Therefore, symmetries like $e^{i\pi Q_m}e^{i\pi Q_w}$ are termed ``emanant symmetries" \cite{Cheng2022}.

In the absence of any defects, the energy spectrum of the theory is easily found to be
\begin{equation}
    E=\frac{\pi v}{L}\left( \frac{Q_w^2}{K}+KQ_m^2\right) + \textrm{oscillators}.
    \label{eq:E}
\end{equation}
 Here ``oscillators" denotes the energies of harmonic oscillator modes, which can be safely ignored for our discussions.

Let us study the defect Hamiltonian $H[\alpha]$ using the IR theory. The continuum limit of Eq. \eqref{eq:Ham-theta-1} is \footnote{The same result can be reached by twisting the boundary condition of the fields by the symmetry $e^{i\alpha Q_m}$. From $e^{i\alpha Q_m}\theta e^{-i\alpha Q_m}=\theta+\alpha$, we impose the boundary condition $\theta(x+L)=\theta(x)+\alpha$ mod $2\pi$. As a result, the eigenvalues of $Q_w$ are modified to $N_w+\frac{\alpha}{2\pi}$, where $N_w\in \bZ$, while the eigenvalues of $Q_m$ are still quantized to integers.}
\begin{equation}
    {\cal H}[\alpha]=\frac{v}{4\pi}\left[K^{-1}\left(\partial_x\vartheta-\frac{\alpha}{L}\right)^2+K(\partial_x\phi)^2\right].
\end{equation}
Hence in the expression Eq. \eqref{eq:E} for $E$, we replace $Q_w$ with $Q_w-\frac{\alpha}{2\pi}$. As $\alpha$ increases from $0$ to $2\pi$, there is a level crossing between $Q_m=0, Q_w=0$ and $Q_m=0, Q_w=1$ at $\alpha=\pi$. These two states have opposite values of $T\sim e^{i\pi Q_m}e^{i\pi Q_w}$. One can also show that at $\alpha=\pi$ these states are mapped to each other under (twisted) charge conjugation, consistent with the discussion in Sec. \ref{subsec: LSM for spin-1/2 chain}.

We have thus shown that the IR theory Eq. \eqref{eq: LL} together with the symmetry implementation reproduces the expected behavior derived from the UV data in Sec. \ref{subsec: LSM for spin-1/2 chain}. This is an example of UV-IR anomaly matching. For a more comprehensive discussion of this example, see Ref. \cite{Cheng2022}.

What other field theories can ``saturate" the anomaly? Let us limit ourselves to conformal field theories. Given the SO(3) symmetry, a natural class of CFTs to consider is the ${\rm SU}(2)_k$ WZW models. Using a similar method, it is shown that $k$ must be odd (even) if ${\rm SU}(2)_k$ is realized as the IR theory of a half-integer (integer) spin chain \cite{Yao2018, ChengPRR2020}.

\subsection{Filling Anomaly} \label{subsec: filling}

The general LSM theorem for spin chains requires a nontrivial projective representation per unit cell. Another way to force an LSM-like constraint is to have ``fractional charge" per unit cell. Again, for simplicity, consider a translation symmetric 1D system with U(1) charge conservation, and we assume that the total charge $Q$ is quantized to integers. For example, this may be a spin-1/2 chain with $\U$ spin rotation symmetry around a fixed axis, where the charge $Q$ is defined as $Q=\sum_i (S_i^z-\frac12)$. The filling fraction $\nu$ is defined as the average amount of charge per unit cell. We are interested in lattice systems with a fixed $\nu$. It is not difficult to generalize the argument in Sec. \ref{subsec: LSM for spin-1/2 chain} to the present case. If we similarly define the defect Hamiltonian $H[\alpha]$, the modified translation operator $T(\alpha)$ satisfies \cite{Cheng2022}
\begin{equation}
    T(\alpha)^{L}=e^{i\alpha Q}, T(\alpha+2\pi)=T(\alpha).
\end{equation}
Here $Q$ is the total charge.
We can attempt to get rid of the phase factor in the first relation by redefining $T(\alpha)\rightarrow e^{i\alpha\frac{Q}{L}} T(\alpha)$, which changes it to $T(\alpha)^L=1$, but then the second relation is modified to $T(\alpha+2\pi)=e^{-2\pi i\nu}T(\alpha)$.
In particular, as long as $\nu$ is not an integer, one finds an obstruction to a non-degenerate gapped ground state similar to the one discussed in Sec. \ref{subsec: LSM for spin-1/2 chain}.

\section{LSM IN HIGHER DIMENSIONS} \label{sec: higher dimensions}

Many of the results discussed in Sec. \ref{sec: 1D} can be generalized to higher dimensions, which we review below. Before diving in, it is worth highlighting some important differences between higher-dimensional systems and 1D systems. First, the crystalline symmetries are much richer in higher dimensions. There are only 2 space groups in 1D, but there are 17 space groups in 2D and 230 in 3D. Second, as discussed in Sec. \ref{sec: 1D}, a 1D system with a LSM anomaly may spontaneously break the anomalous symmetry or be gapless. In higher dimensions, new possibilities arise, e.g., it can also be a symmetric gapped topological phase. In this section, we focus on the LSM constraints themselves. In Sec. \ref{sec: anomaly matching}, we review the general framework for anomaly matching and demonstrate it using various higher-dimensional examples.

In the simplest case where the symmetry is $\so(3)\times\bZ^d$, with $\so(3)$ the spin rotational symmetry and $\bZ^d$ the $d$-dimensional translation symmetry, the most general rigorous result known to date is the following theorem.

\begin{theorem}
    With the symmetry setting above, if each unit cell contains a spin-1/2 moment and if the interactions in the Hamiltonian decay as $1/r^a$ with $a>4d-2$ ($r$ is the distance between the interacting spins), the Hamiltonian cannot have a unique gapped ground state.
\end{theorem}
The original version of this theorem for short-range Hamiltonians was proved by Hastings in Ref. \cite{HastingsLSM}. The extension to long-range interactions was established recently in Ref. \cite{Ma2024}.

In other symmetry settings, where the on-site symmetry may be discrete and the crystalline symmetry may go beyond translation, rigorous results are rare, but an important picture known as ``lattice homotopy" is very powerful \cite{PoPRL2017, Else_LSM_2019}. In essence, given a symmetry group, lattice homotopy is a way to organize the infinitely many types of lattice systems into a finite set of equivalence classes, such that two lattice systems have different (the same) anomalies if they are in different (the same) classes. The simplest setup to demonstrate the notion of lattice homotopy is lattice systems with symmetry $G_{\rm s}\times G_{\rm int}$, where $G_{\rm s}$ is a space group symmetry and $G_{\rm int}$ is an on-site symmetry. Then two lattice systems are in the same lattice homotopy class if and only if they can be deformed into each other by these operations: 1) moving the microscopic DOF while preserving $G_{\rm s}$, 2) at each location, identifying DOF with the same type of projective representation under $G_{\rm int}$, in a way preserving $G_{\rm s}$.

For example, consider a spin-1/2 system on a honeycomb lattice with $G_{\rm s}=p6m$ and $G_{\rm int}={\rm SO}(3)$. By deformation 2), we can view each site as hosting 3 spin-1/2 moments. By deformation 1), each of these 3 spin-1/2 moments can be moved along one of the three edges of the honeycomb lattice that connect to this site. When they arrive at the middle of the edges, they form a kagome lattice with each site hosting 2 spin-1/2 moments. Finally, by deformation 2), this kagome lattice system is identified with one where all DOF are spin singlets. Therefore, according to lattice homotopy, the honeycomb lattice spin-1/2 system with $p6m\times \so(3)$ symmetry is anomaly free. Indeed, a $p6m\times \so(3)$ symmetric SRE state exists on the spin-1/2 honeycomb lattice \cite{FeaturelessHoneycomb}.

More generally, one can see that all lattice systems with $p6m\times \so(3)$ symmetry are classified into 4 types by lattice homotopy, with representative systems being a lattice with only spin-singlet DOF, a triangular lattice spin-1/2 system, a kagome lattice spin-1/2 system, and a system with spin-1/2 moments at both the triangular and kagome sites. The first among the four is anomaly free, and the other three all have different LSM anomalies. The anomaly-free systems allow a symmetric SRE state and symmetric local gapped Hamiltonians with a unique ground state, while the anomalous systems do not. We remark that the last among the four types has an even number of spin-1/2 moments per unit cell, so with only translation and $\so(3)$ symmetries it is anomaly free, and the anomaly is actually a mixed anomaly between the $\so(3)$ and lattice $C_2$ rotation symmetries.

\section{ANOMALY MATCHING} \label{sec: anomaly matching}

\subsection{General Theory of UV-IR Matching} \label{subsec: general anomaly matching}

It is useful to phrase UV-IR anomaly matching in more general, albeit slightly abstract, terms. First, we clarify the meaning of ``UV" and ``IR". Throughout this section, UV refers to the lattice system, while IR means a continuum field theory valid at energy scales far below the UV scale. This IR theory does not need to be the one at the extreme IR; it may be an intermediate-scale ``mid-IR" effective theory, which will still be referred to as an IR theory for notational simplicity.

An IR theory can be characterized by its symmetries. We are mostly interested in its internal symmetries, as the (continuous) spacetime symmetries in the continuum theories do not play much role in anomaly matching. Notably, it is quite common that the IR theory embodies internal symmetries not present in the UV. We have seen such an example in the spin-1/2 Heisenberg chain, where the deep IR theory (the SU(2)$_1$ CFT) has SO(4) symmetry, while the UV lattice model only has SO(3) internal symmetry. Symmetries that are present in the IR theory but absent in the UV are called ``emergent". From the perspective of renormalization group (RG), emergent symmetries are broken by IR irrelevant perturbations.

Denote the symmetry of the UV system by $G_{\rm UV}$. Suppose the IR theory has (emergent) symmetry $G_{\rm IR}$.  Since the UV and IR theories are connected by the RG flow, all UV symmetries must have their counterparts in the IR\footnote{It is possible that a UV symmetry acts trivially in the IR theory.}, which, mathematically, are captured by a group homomorphism $\varphi$ between them:
\begin{equation} \label{eq: symmetry embedding}
    \varphi: G_{\rm UV}\rightarrow G_{\rm IR}.
\end{equation}
This mapping $\varphi$ describes the ``symmetry enrichment" of the IR theory. Note that given the same $G_{\rm UV}$ and $G_{\rm IR}$ (and the same IR theory), there can be multiple choices of $\varphi$, which then correspond to physically distinct ``symmetry-enriched" phases.

Besides the group homomorphism, to examine whether an IR theory is valid for a given UV system, we also need to account for the 't Hooft anomaly, which describes, roughly speaking, certain topological responses of the system to defects, as reviewed in Sec. \ref{subsec: LSM for spin-1/2 chain} for the LSM anomaly in spin chains. The anomaly is independent of energy scale and remains invariant under RG flows, so the IR theory must produce the same response, which was demonstrated in a concrete example in Sec. \ref{subsec: anomaly matching 1d}. Here we will present a more general and compact formalism.

First, suppose the UV system is coupled to background gauge field $A_{\rm UV}$ of $G_{\rm UV}$. The homomorphism $\varphi$ means that the background gauge field $A_{\rm IR}$ of $G_{\rm IR}$ can be expressed as
\begin{equation}
  A_{\rm IR}=\varphi(A_{\rm UV}).
\end{equation}

In order to discuss the anomaly matching condition, we assume that the anomaly is abstractly described by a ``function" $\omega_{\rm UV}$ of $A_{\rm UV}$, which gives the response of the system when the background field $A_{\rm UV}$ is turned on\footnote{If the system is in the continuum and Lorentz-invariant, then $\omega$ is more conveniently presented as a response action $S_{\rm UV}[A_{\rm UV}]$ in one dimension higher.}.  The two anomalies $\omega_{\rm UV}[A_{\rm UV}]$ and $\omega_{\rm IR}[A_{\rm IR}]$ are then related as:
\begin{equation} \label{eq: anomaly matching condition}
    \omega_{\rm UV}[A_{\rm UV}]=\omega_{\rm IR}[A_{\rm IR}]=\omega_{\rm IR}\big[\varphi(A_{\rm UV})\big].
\end{equation}
We say that the UV anomaly is a ``pull-back" of the IR anomaly, and write the above relation as $\omega_{\rm UV}=\varphi^*(\omega_{\rm IR})$, which is the anomaly matching condition.

A simple but nontrivial observation from Eq. \eqref{eq: anomaly matching condition} is that in any microscopic system with a nontrivial anomaly, i.e., $\omega_{\rm UV}\neq 0$, the low-energy theory cannot be a conventional Wilson-Fisher theory, because such Wilson-Fisher theories allow a trivially disordered phase and thus have $\omega_{\rm IR}=0$, which cannot satisfy Eq. \eqref{eq: anomaly matching condition}. However, some ``fractionalized" analog of Wilson-Fisher theories can still match the LSM anomalies, and they were indeed identified in various anomalous systems \cite{Chubukov1993, Chubukov1994}.

Two remarks are in order.
\begin{enumerate}

    \item Coupling to background gauge fields for internal symmetries is standard. However, it is much less clear what ``gauge fields of spatial symmetries" mean. Here we apply the following heuristics: When dealing with gauge fields of a spatial symmetry described by a group $G$, one can simply treat them as gauge fields of some internal $G$-symmetry, while keeping in mind that any unitary orientation-reversing symmetry should be viewed as an anti-unitary internal symmetry in this mapping. We note that for the related problem of classifying gapped phases with crystalline symmetry, this heuristic idea has been more formally established as the ``crystalline equivalence principle" in Ref. \cite{ThorngrenPRX2018}.

    \item  So far, all symmetries under consideration act on the entire system. In other words, charged objects are point-like objects, i.e., particles.  However, this is not the whole picture in many IR theories. The need to extend the notion of symmetries is best illustrated when the IR theory is a topological quantum field theory (TQFT). In general, TQFTs can only have finite global symmetry that acts faithfully (in the sense defined so far), and naively the matching condition cannot be satisfied. The resolution is that one has to consider a more general notion of symmetries, whose charged objects are not point-like, but are extended in spacetime (e.g., loops in 2+1d). These ``generalized" symmetries are called higher-form symmetries \cite{Gaiotto:2014kfa}. However, in Sec. \ref{subsec: SETQSL}, we will give another approach for anomaly matching in a topological IR theory without using the language of higher-form symmetries.

\end{enumerate}

\subsection{LSM Anomalies for Various Lattice Systems} \label{subsec: LSM for various lattice systems}

In order to carry out the above anomaly matching analysis, we need to have mathematical descriptions of the UV anomaly $\omega_{\rm UV}$ and the IR anomaly $\omega_{\rm IR}$ (or its pullback $\varphi^*(\omega_{\rm IR})$). In this subsection, we focus on the UV anomaly $\omega_{\rm UV}$ for various lattice systems, and the next two subsections deal with the anomaly matching for specific examples of IR theories.

The UV anomaly, in its simplest case that only involves the lattice translation symmetry and another on-site symmetry, was derived in Ref. \cite{ChengPRX2016}. The generalizations to include an arbitrary space group symmetry were made in Ref. \cite{Ye2021} for 1D and 2D systems and in Ref. \cite{Liu2024} for 3D systems. Concretely, for any lattice system in $d=1,2,3$ spatial dimensions with symmetry group $G_{\rm UV}=G_{\rm s}\times G_{\rm int}$, where $G_{\rm s}$ is an arbitrary space group and $G_{\rm int}$ is an on-site symmetry whose projective representations are classified by $\mathbb{Z}_2^k$ (i.e., there exists a positive integer $k$ such that $H^2(G_{\rm int}, \U)=\mathbb{Z}_2^k$), then $\omega_{\rm UV}$ can be expressed as an element in the group cohomology $H^{d+1}(G_{\rm UV}, \U)$:
\begin{equation} \label{eq: LSM anomaly cocycle}
    \omega_{\rm UV}(g_1, g_2, \cdots, g_{d+1})=e^{i\pi (\lambda\cup \eta)(g_1,g_2,\cdots, g_{d+1})}=e^{i\pi \lambda(l_1, l_2, \cdots, l_{d-1})\eta(a_d, a_{d+1})}
\end{equation}
where $g_i\in G_{\rm UV}$, $l_i\in G_{\rm s}$ and $a_i\in G_{\rm int}$, such that $g_i=l_i\otimes a_i$, for $i=1,2,\cdots, d+1$. In the above, $e^{i\pi\eta(a_d, a_{d+1})}$ represents an element in the group cohomology $H^2(G_{\rm int}, \U)$, which physically means the projective representations carried by the DOF in the system. The other factor, $\lambda(l_1, l_2, \cdots, l_{d-1})$, represents an element in the group cohomology $H^{d-1}(G_{\rm s}, \mathbb{Z}_2)$, which physically describes the type of lattice under consideration. Intuitively, $g_i$ here can be viewed as the gauge field $A_{\rm UV}$ in Sec. \ref{subsec: general anomaly matching}, and $\omega_{\rm UV}$ above describes how the system responds to the gauge field.

A few remarks are in order. First, the symmetry settings above cover a large class of lattice systems relevant to experimental and numerical studies, especially in the context of quantum spin liquids, where $G_{\rm int}$ can be the $\so(3)$ spin rotation, $\mathbb{Z}_2^\mathsf{T}$ time reversal, etc. Second, we note that Eq. \eqref{eq: LSM anomaly cocycle} takes a notably simple factorized form, where the information about the type of lattice is completely encoded in $\lambda$, and the information about the projective representation of the DOF is completely encoded in $\eta$. Moreover, given the type of lattice, $\lambda$ has a universal expression for all $G_{\rm int}$ and all $\eta$. We also note that not all elements in $H^{d-1}(G_{\rm s}, \mathbb{Z}_2)$ can be a valid $\lambda$, and an important contribution of Refs. \cite{Ye2021, Liu2024} is to determine the $\lambda$ for each lattice.

\begin{figure}[h]
    \centering
    \includegraphics[width=0.6\textwidth]{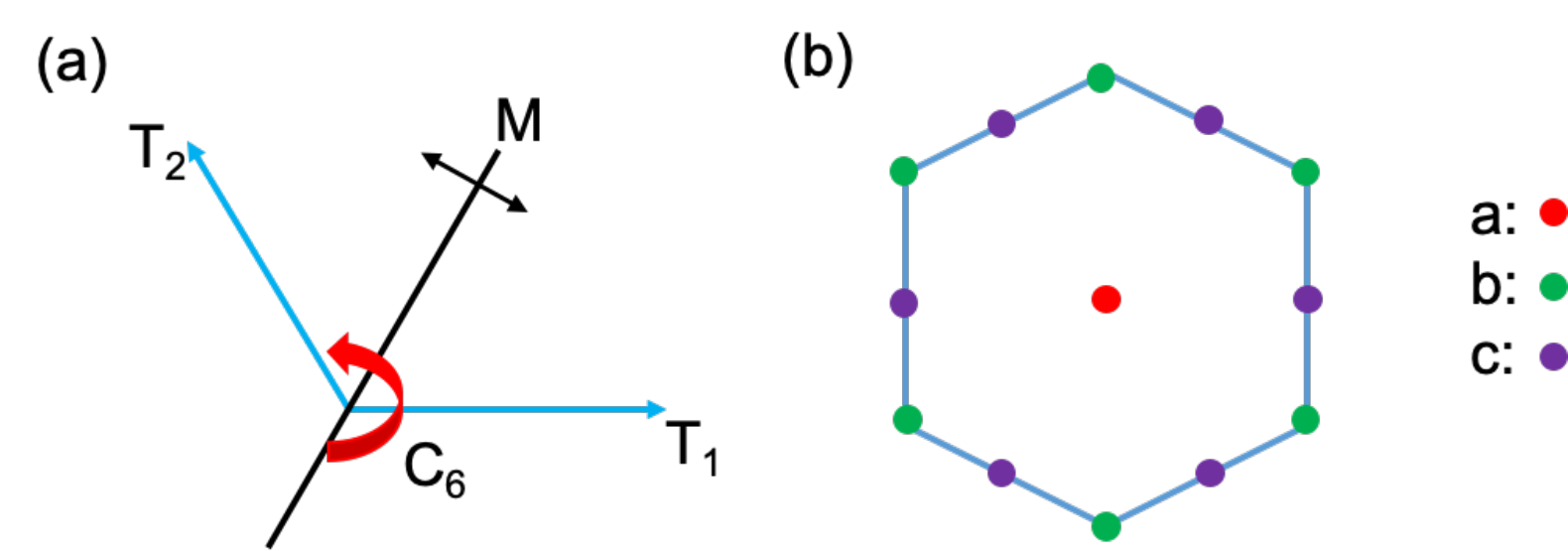}
    \caption{Panel (a) shows the generators of the $p6m$ group, including translations $T_1$ and $T_2$, a 6-fold rotation $C_6$ and a mirror reflection $M$. The two translation vectors have the same length, and their angle is $2\pi/3$. The reflection axis of $M$ bisects these two translation vectors and its angle with both translation vectors is $\pi/3$. In panel (b), the hexagon is a translation unit cell of the $p6m$ lattice symmetry. There are three irreducible Wyckoff positions, labelled by $a$, $b$ and $c$, and they form the sites of the triangular, honeycomb and kagome lattices, respectively. The $C_6$ rotation center in panel (a) is at a type-$a$ point.}
    \label{fig:p6m}
\end{figure}

Below we give some intuition about how $\lambda$ is obtained, using the 2D example where $G_s=p6m$, the space group symmetry shared by the triangular, kagome and honeycomb lattices. Because $\lambda\in H^2(p6m, \mathbb{Z}_2)$ should depend only on the type of the lattice, but not $G_{\rm int}$, we assume that $G_{\rm int}$ is $\so(3)$ spin rotation. Then one can imagine coupling the system to a background $\so(3)$ gauge field. An $\so(3)$ gauge field has a type of defects known as $\so(3)$ monopole, which is the $\so(3)$ analog of the more familiar Dirac monopole in a $\U$ gauge field. An important difference between these two types of monopoles is that the combination of two $\so(3)$ monopoles is gauge-equivalent to no $\so(3)$ monopole, i.e., they have a $\bZ_2$ fusion rule.

Now we can ask how the $p6m$ symmetry acts on the $\so(3)$ monopole, and the different symmetry actions are precisely classified by $H^2(p6m, \mathbb{Z}_2)$ \cite{Barkeshli2014}. For example, although the two translations $T_{1,2}$ in \textbf{Figure \ref{fig:p6m}} should commute on local operators, their actions on the $SO(3)$ monopole may anti-commute, signaling a 't Hooft anomaly associated with the translation and $\so(3)$ symmetries. To see this, consider the effect of $T_2^{-1}T_1^{-1}T_2T_1$ on the $SO(3)$ monopole. This operation moves this monopole around a unit cell. Suppose the unit cell includes a half-integer (integer) spin. This process is analogous to moving a $2\pi$ magnetic flux around a half-integer (integer) charge, so a $-1$ (1) Aharonov-Bohm phase factor will be generated and $T_{1,2}$ anti-commute (commute) on the monopole. So the spin moments in a unit cell dictate the mixed anomaly between the translation and $SO(3)$ symmetries. Continuing this line of reasoning, one can see that the type of the lattice, or more precisely, the position and the spin quantum number of the magnetic moments, completely determines how the entire $p6m$ symmetry acts on an $\so(3)$ monopole and therefore fixes an element in $H^2(p6m, \mathbb{Z}_2)$. This element is precisely $\lambda$. On the other hand, the spin quantum numbers determine $\eta$. Altogether, $\lambda$ and $\eta$ determine the UV anomaly via Eq. \eqref{eq: LSM anomaly cocycle}.

Given a symmetry setting, namely, given the symmetry group $G_{\rm UV}$, the type of lattice, and the projective representations of the DOF, one can find the explicit expressions of $\omega_{\rm UV}$ in Refs. \cite{ChengPRX2016, Ye2021, Liu2024}. To perform the anomaly matching analysis, one then just needs the IR anomaly $\omega_{\rm IR}$ (or its pullback $\varphi^*(\omega_{\rm IR})$). In the following subsections, we review concrete examples.

\subsection{Dirac Spin Liquids}

The first example is the Dirac spin liquid (DSL), which was originally proposed in the context of high-temperature superconductors \cite{Affleck1988, Wen1995}. Recently, evidence of the emergence of DSL has been reported in kagome and triangular antiferromagnets \cite{Ran2006, Iqbal2016, He2016, Ranjith2019, Ding2019, Bordelon2019, Hu2019, Wietek2023, Bag2023}. An intuitive way to think of the DSL is to ``fractionalize" each localized spin into two fermions, and these fermions experience some band structure that has 4 Dirac cones (including all spins and valleys). As is common in fractionalization phenomena, these fermions are coupled to an emergent dynamical gauge field, which, in this case, is a $\U$ gauge field. At low energies, there is evidence that the DSL is described by an emergent CFT \cite{Karthik2015, Karthik2016, He2021}.

For our purpose, it is more appropriate to use the gauge-invariant monopole operators as the fundamental low-energy DOF of the DSL, instead of the Dirac fermions because the latter is not gauge invariant.  From the monopole operator (and its conjugate variable, which is the field strength of the $\U$ gauge field) all gauge invariant operators can be constructed. The monopole operators in this DSL can be organized into a 6-by-2 matrix field $n$ with Hermitian entries \cite{Zou2021}.

The IR symmetry group of the DSL, apart from the Poincar\'e symmetry\footnote{Assuming that the DSL is indeed a CFT, then the full spacetime symmetry should be the conformal group.}, is
\begin{equation}
    G_{\rm IR}=\frac{\mathrm{O}(6)^\mathsf{T}\times \mathrm{O}(2)^\mathsf{T}}{\mathbb{Z}_2}
\end{equation}
Here the superscripts ``$\mathsf{T}$" indicate that all spacetime orientation-reversing symmetries in this theory must be combined with an improper $\mathbb{Z}_2$ operation in either the ${\rm O}(6)$ or ${\rm O}(2)$ group. The quotient of $\mathbb{Z}_2$ indicates that the center of the ${\rm O}(6)$ symmetry and the $\pi$-rotation of the $\so(2)\subseteq {\rm O}(2)$ symmetry are identified. In $G_{\rm IR}$, ${\rm O}(6)$ (${\rm O}(2)$) acts on the matrix $n$ by left (right) multiplication. The IR anomaly, $\omega_{\rm IR}$ in Sec. \ref{subsec: general anomaly matching}, for the DSL is given in Eq. (21) of Ref. \cite{Ye2021}.

Given a lattice system with symmetry $G_{\rm UV}$ and anomaly $\omega_{\rm UV}$, if the DSL emerges as its IR theory, the symmetry-enrichment pattern $\varphi$ in Eq. \eqref{eq: symmetry embedding}, which is characterized by how the matrix $n$ transforms under $G_{\rm UV}$, must satisfy the anomaly matching condition Eq. \eqref{eq: anomaly matching condition}. We have thus a well-defined mathematical problem of classifying all possible symmetry-enrichment patterns. This classification has been performed for common lattice systems in Ref. \cite{Ye2021}. For example, on triangular lattice spin-1/2 systems with $G_{\rm UV}=p6m\times \so(3)\times\mathbb{Z}_2^\mathsf{T}$, where good evidence for an emergent DSL has been reported in various models and materials \cite{Iqbal2016, Ranjith2019, Ding2019, Bordelon2019, Hu2019, Wietek2023, Bag2023}, there are three possible $\varphi$'s that are ``stable", i.e., all $G_{\rm UV}$-symmetric relevant perturbations are forbidden. For the models and materials cited above, $\varphi$ is given by
\begin{equation} \label{eq: DSL standard}
\begin{split}
&T_1: n\rightarrow\left(
\begin{array}{cccc}
I_3 & & &\\
& 1 & &  \\
& & -1 &  \\
& & & -1  \\
\end{array}
\right)n
\left(
\begin{array}{cc}
    -\frac{1}{2} & -\frac{\sqrt{3}}{2}\\
    \frac{\sqrt{3}}{2} &-\frac{1}{2}
\end{array}
\right),
\quad
T_2: n\rightarrow\left(
\begin{array}{cccc}
I_3 & & &\\
&-1 & & \\
& & 1 & \\
& & & -1 \\
\end{array}
\right)n
\left(
\begin{array}{cc}
    -\frac{1}{2} & -\frac{\sqrt{3}}{2}\\
    \frac{\sqrt{3}}{2} &-\frac{1}{2}
\end{array}
\right),\\
&C_6: n\rightarrow
\left(
\begin{array}{cccc}
I_3 & & & \\
& & 1 &  \\
& & & 1 \\
& -1 & & \\
\end{array}
\right)n
\left(
\begin{array}{cc}
1& \\
 &-1\\
\end{array}
\right),
\quad
M: n\rightarrow
\left(
\begin{array}{cccc}
I_3 & & &\\
& & -1 & \\
&-1 & & \\
& & & 1 \\
\end{array}
\right)n,\\
&{\rm O}(3)^\mathsf{T}: n\rightarrow
\left(
\begin{array}{cc}
{\rm O}(3)^\mathsf{T} & \\
 I_3  \\
\end{array}
\right)n,
\end{split}
\end{equation}
where $T_{1,2}$ are the generators of the translation symmetries, $C_6$ is the generator of the 6-fold lattice rotation symmetry, $M$ is the generator of the reflection symmetry (see \textbf{Figure \ref{fig:p6m}}), and ${\rm O}(3)^\mathsf{T}=\so(3)\times\mathbb{Z}_2^\mathsf{T}$ is the combination of $\so(3)$ spin rotation and $\mathbb{Z}_2^\mathsf{T}$ time reversal symmetries. These symmetry actions indicate what types of magnetic order and dimer order can be obtained when a DSL undergoes a phase transition.

Besides the above symmetry-enrichment pattern that has been previously discovered, the anomaly-based classification in Ref. \cite{Ye2021} predicted various exotic symmetry-enrichment patterns of DSL that were previously unknown, as well as symmetry-enrichment patterns of some even more exotic ``non-Lagrangian" quantum matter \cite{Zou2021}. It still remains to find concrete models to realize these predicted phases.

\subsection{Topological Quantum Spin Liquids} \label{subsec: SETQSL}

The next example is 2D symmetry-enriched topological quantum spin liquids (TQSLs). A TQSL is a gapped quantum phase of matter in a spin system, which hosts emergent anyons, i.e., quasi-particles that are neither bosons nor fermions and are characterized by more exotic braiding and fusion properties. TQSLs with different sets of anyons are different phases. In the presence of symmetries, TQSLs with the same set of anyons can also be different phases with different symmetry-enrichment patterns \cite{Zeng2015}.

For instance, a $\mathbb{Z}_2$ TQSL has 4 types of anyons, often denoted by 1, $e$, $m$ and $\epsilon$, where 1 is a trivial anyon that is similar to the conventional magnons, $e$ and $m$ are both anyons with bosonic self-statistics but semionic mutual braiding statistics{\footnote{That two particles have semionic mutual braiding statistics means that when one particle moves around the other by a full circle, the many-body state will get a $-1$ phase factor.}}, and $\epsilon$ can be viewed as the bound state of $e$ and $m$, which has fermionic self-statistics and semionic mutual braiding statistics with both $e$ and $m$ \cite{Kitaev1997}. If the $\mathbb{Z}_2$ TQSL further has an $\so(3)$ spin rotational symmetry, then nontrivial anyons $e$, $m$ and $\epsilon$ may carry half-integer-spin projective representations of the $\so(3)$ symmetry, instead of only carrying integer-spin linear representations as do magnons. Anyons carrying half-integer spins is an example of symmetry fractionalization, and different patterns of symmetry fractionalization signal different symmetry-enriched $\mathbb{Z}_2$ TQSLs \cite{EssinPRB2013, Barkeshli2014}.

Although evidence for TQSLs in solid-state materials remains inconclusive, they have been realized in small-scale quantum simulators based on Rydberg atoms, trapped ions and superconducting qubits \cite{Satzinger2021, Lukin2021, Iqbal2023, Foss-Feig2023}. A pertinent question is: Given the symmetry settings of a system, which symmetry-enriched TQSLs are compatible with them?

Below we discuss a concrete example, i.e., a $\bZ_2$ TQSL in triangular-lattice spin-1/2 systems with $G_{\rm UV}=p6m\times \so(3)\times\bZ_2^\mathsf{T}$, first worked out in Refs. \cite{Qi2015, Qi2016}. In this context, we only need the fractionalization patterns of the $G_{\rm UV}$ symmetry on the $e$ and $m$ anyons, which can be largely understood using general arguments, without doing complicated calculations.

First, the anomaly purely associated with the on-site $\so(3)\times\bZ_2^\mathsf{T}$ symmetry vanishes, which leads to the following fractionalization patterns of $\so(3)\times \bZ_2^\mathsf{T}$: One of $e$ and $m$ must carry no fractional quantum number, i.e., it transforms as integer spin under $\so(3)$ and singlet under $\bZ_2^\mathsf{T}$. All other fractionalization patterns are anomalous \cite{Vishwanath2013, Wang2013, Zou2016, NingPRR2020}. Without loss of generality, we let $m$ carry no fractional quantum number. Then the mixed anomaly between $\so(3)\times \bZ_2^\mathsf{T}$ and $p6m$, the LSM anomaly, implies that $e$ carries half-integer-spin and is a Kramers doublet, i.e., it is a ``spinon".

We can further fix the fractionalization pattern of the $p6m$ symmetry on $m$ completely. To this end, imagine inserting an $\so(3)$ monopole into the system, which can be viewed as a $2\pi$ flux of the $\U$ spin rotation symmetry around the $z$ axis. As the $2\pi$ flux is gauge-equivalent to no flux, this process nucleates an anyon $v$. The fact that $e$ ($m$) carries half-integer (integer) spin means that $e$ ($m$) experiences a $-1$ (1) Aharonov-Bohm phase factor when circling around this flux, which uniquely fixes the braiding phase between $e$ ($m$) and $v$, and further fixes $v=m$. This argument is essentially Laughlin's flux-threading argument in the fractional quantum Hall effect.

Finally, note that the discussion in Sec. \ref{subsec: LSM for various lattice systems} shows that the projective representation of the $p6m$ symmetry carried by the $\so(3)$ monopole is completely fixed by the type of the lattice. Therefore, the projective representation under the $p6m$ symmetry carried by $m$ is also completely fixed by the type of the lattice, which can be read off from Sec. II B 1 in Ref. \cite{Ye2021}. For example, for $m$, translations along the two directions do not commute.

Taken together, the above discussion shows that the symmetry fractionalization pattern of the entire $G_{\rm UV}=p6m\times \so(3)\times\bZ_2^\mathsf{T}$ symmetry on $m$ is fully fixed, and $e$ carries half-integer spin under $\so(3)$ and Kramers doublet under $\bZ_2^\mathsf{T}$. However, the fractionalization pattern of the $p6m$ symmetry on $e$ has not been determined. To determine it, one way to proceed is to perform the systematic anomaly-matching analysis in Ref. \cite{Ye2023}. A general framework to calculate the pullback of the IR anomaly $\varphi^*(\omega_{\rm IR})$ of a symmetry-enriched TQSL is given in Ref. \cite{Ye2022}. Using the UV anomaly obtained in Ref. \cite{Ye2021}, one can then classify all possible symmetry-enrichment patterns compatible with the anomaly-matching condition Eq. \eqref{eq: anomaly matching condition}. We refer readers to these references for more details, as they are relatively technical.  Alternatively, one can also derive the symmetry fractionalization pattern of $e$ using the physical picture that the background anyon ($e$ here) must exhibit the same fractionalization pattern of space group as the DOF in the unit cell (see Ref. \cite{Qi2016}).

\subsection{Compressible Phases} \label{subsec: compressible}

As our last example, we briefly review the matching of filling anomaly in compressible phases. Here we consider a system with $\U$ and translation symmetries, and a compressible phase is one where the low-energy effective field theory of the system takes the same form at different filling fractions. The most familiar examples are superfluids and Fermi liquids (FLs).
Being compressible is a rather strong condition. For example, Dirac and Weyl semimetals are not compressible in this sense, because when the filling fraction changes, they will become FLs and the effective theory changes.

By matching the filling anomaly, Ref. \cite{Else2020} pointed out that, in spatial dimensions larger than 1, the internal symmetry of the effective theory of a compressible phase cannot be a 0-form symmetry described by a finite-dimensional compact Lie group. To get some intuition of this result, note that the effective theory of a $(d+1)$-dimensional superfluid actually has a $(d-1)$-form symmetry, which corresponds to the conservation of vorticity \cite{Gaiotto:2014kfa, Delacretz2019}. In FLs, the symmetry of the effective theory is a ``loop group" of the $\U$, which corresponds to the conservation of fermion number at each point on the Fermi surface \cite{Else2020}. In both cases, the symmetries of effective theories are not 0-form symmetries described by a finite-dimensional compact Lie group. We refer the readers to Refs. \cite{Else2020, Else2025} for more details. The above result leads to multiple further applications. For example, it can be used to show the Luttinger theorem \cite{Else2020}, which was first obtained perturbatively \cite{Luttinger1960} and later proved using non-perturbative topological arguments \cite{Oshikawa2000}.

A more exotic type of compressible phase is the fractionalized Fermi liquids (FL$^*$) \cite{Senthil2002, Senthil2003}, which can be viewed as an FL stacked on top of a TQSL discussed in Sec. \ref{subsec: SETQSL}. The anomaly of an FL$^*$ can be viewed as the ``sum" of those of an FL and a TQSL, and the anomaly matching condition imposes strong constraints on how the Luttinger theorem can be violated and which types of FL$^*$ can emerge at a given filling factor \cite{Bonderson2016, Bonetti2026}.

\section{OTHER GENERALIZATIONS} \label{sec: other generalizations}

In this section, we discuss various other generalizations of the LSM anomalies.  There are also generalizations not covered in this review due to length constraints, and we refer the readers to Refs. \cite{Kobayashi2018, Pace2024, Pace2025, Pace2026, Anakru2026, Ning2026} for details.

\subsection{Disordered LSM and Average Anomaly}

In real solids, lattice symmetries are always broken by impurities. However, viewed as an ensemble (assuming self-averaging), lattice symmetries like translations are usually restored ``on average". That is, while the symmetry is completely broken for any given disorder realization, in the ensemble any two disorder realizations related to each other via translations occur with the same probability.  Therefore, it is important to understand to what extent the LSM constraints apply in this case. Remarkably, LSM anomalies can still be robust if the lattice symmetries are only preserved on average \cite{Kimchi2017, Ma2022, Ma2023, Panahi2026}.

As a concrete example, consider a spin chain with an average translation symmetry, which still allows one to define a unit cell. Moreover, suppose there is an on-site $\so(3)$ spin rotational symmetry such that each unit cell contains a spin-1/2 moment. If the translation symmetry was exact, in Sec. \ref{sec: 1D} we see that this symmetry setting would lead to a nontrivial LSM anomaly that forces all symmetric states to be LRE. Interestingly, when the translation symmetry is only preserved on average in an ensemble of such systems{\footnote{By the Imry-Ma argument, such an average translation symmetry cannot be broken spontaneously in 1D.}}, then the probability for $\so(3)$ symmetric ground states to be LRE approaches 1 in the thermodynamic limit \cite{Ma2022}. That is, the ensemble is LRE. One can further establish an analog of Theorem \ref{thm: dichotomy} that bounds the decay of the ensemble-averaged spin correlation function \cite{Kimchi2017}.

\subsection{Fermionic Systems}

Up to this point we have focused on spin systems, which are ``bosonic". In this subsection, we discuss briefly extensions to fermionic systems.

A physically clear example is spin-1/2 electrons at half filling (i.e., one electron per unit cell). Suppose there is a large Hubbard-like interaction suppressing double occupancy. Then the effective description of the system is a spin-1/2 system satisfying the previously discussed LSM theorem. It is then natural to expect that a similar constraint applies even when the single-occupancy constraint is not strictly enforced. That is, the system cannot be in a symmetric SRE state at half filling.

It is also possible for fermionic systems to exhibit ``intrinsically fermionic" LSM anomalies, which cannot be effectively reduced to one for spin systems. One such example is a translation invariant lattice with one Majorana fermion per unit cell, i.e., each unit cell $i$ contains a Majorana operator $\gamma_i$, which satisfies $\{\gamma_i, \gamma_j\}=2\delta_{ij}$. One can prove that such a system does not admit any translation symmetric SRE state \cite{HsiehPRL2016}. A systematic discussion of these fermionic LSM theorems can be found in Refs. \cite{Cheng2018} and \cite{OmerPRB2021}.

\subsection{SPT-LSM Theorems}

So far, all our LSM theorems forbid symmetric SRE states, which is our working definition of anomaly. An interesting extension is the so-called ``SPT-LSM theorem". That is, even though the anomaly is absent, any symmetric SRE ground state must be in a nontrivial phase (possibly protected by the internal symmetry).

For example, consider a 2D lattice of spinless fermions with filling fraction $\nu=\frac{p}{q}$, where $p,q$ are coprime integers. As reviewed in Sec. \ref{subsec: filling}, when $q>1$ there is a filling anomaly. Now suppose the system is placed in a uniform magnetic field, such that the magnetic flux in a unit cell is $\phi$ (in units of $\frac{h}{e}$). If $q\phi\in \bZ$, there is no LSM anomaly since the filling with respect to the magnetic unit cell (which is $q\nu$) is an integer, thus the ground state can be gapped and SRE. Suppose this is the case, then the Hall conductivity $\sigma_{xy}$ should be an integer. It is shown in Ref. \cite{LuRanOshikawa2020} that $\sigma_{xy}$ satisfies the following relation:
\begin{equation}
    \sigma_{xy}\phi = \nu \mod 1.
\end{equation}
It is easy to see that unless $\nu\in \bZ$, the above condition forbids $\sigma_{xy}=0$. Thus any SRE ground state has to have a nonzero Hall conductivity, and it cannot be a completely trivial insulator.

Other examples of SPT-LSM theorems are considered in Refs. \cite{Lu2024, WuHoLu2017, XuSPTLSM}. For the general theory of SPT-LSM theorems, we refer the readers to Refs. \cite{JiangSPTLSM, Else_LSM_2019}.

\section{CRYSTALLINE SYMMETRY-PROTECTED TOPOLOGICAL PHASES} \label{sec: crystalline SPTs}

Finally, we discuss a fruitful approach to characterize 't Hooft anomalies, the ``anomaly inflow". In this approach, the $d$-dimensional system with the anomalous symmetry is viewed as the boundary of a $(d+1)$-dimensional bulk, such that the whole system has no anomaly. That is, the non-gauge-invariant response of the boundary system to the background gauge field is cancelled by the response of the bulk. To achieve this, the bulk has to be a nontrivial SPT phase. This picture can be made precise in the QFT framework, establishing a one-to-one correspondence between 't Hooft anomalies in $d$ dimensions and SPT phases in $(d+1)$ dimensions.

This anomaly inflow picture can also be applied to LSM anomalies, where the bulk is an SPT phase protected by internal and/or crystalline symmetries (hence ``crystalline SPTs") \cite{ChengPRX2016, HuangPRB2017}. As a concrete example, let us revisit the LSM anomaly for the spin-1/2 chain discussed in Sec. \ref{subsec: LSM for spin-1/2 chain}. The 2D bulk is easily seen to be a decoupled stack of Haldane chains (i.e., the 1D nontrivial SPT phase protected by $\so(3)$ symmetry), each, say, along the $y$ direction. Moreover, these Haldane chains have a translation symmetry along the $x$ direction perpendicular to $y$. The spin-1/2 chain emerges at the edge along the $x$ direction. Similar constructions apply to other LSM anomalies discussed in Sec. \ref{sec: higher dimensions}, as well as those reviewed in Sec. \ref{sec: other generalizations}. In all these examples, the anomalies are determined by the nature of the degrees of freedom and the symmetry actions on them, which must be compatible with the bulk properties. However, the filling anomalies in Sec. \ref{subsec: filling} and Sec. \ref{subsec: compressible}, which are due to the $\U$ and translation symmetries, also depend on the eigenspace of the symmetry operator in which the system lies. For example, although the filling anomaly exists in subspaces with fractional filling factors, it disappears in subspaces with integer filling factors. In contrast, Theorem \ref{thm: general LSM for states} applies to all symmetric states. For this reason, the filling anomalies do not admit a standard anomaly-inflow picture. In fact, there is no nontrivial SPT phase protected by both $\U$ and translation symmetries in any dimension \cite{Else2020}.

It is, however, important to emphasize that the bulk here is fictitious and a spin-1/2 chain is perfectly well-defined without the bulk. One may view the bulk as merely a mathematical device to characterize the anomaly. On the other hand, it is also certainly possible to study the boundary of an actual crystalline SPT state. Although the physical contexts are different, the anomaly has the same implications for low-energy physics.

\section{DISCUSSION}

In this review, we have discussed the LSM anomalies as symmetry constraints on the correlation, entanglement and dynamics of a quantum many-body system, and demonstrated how anomaly matching can be used to examine whether a low-energy theory can emerge in a microscopic system. Below we list some important open problems for future study.

\begin{itemize}

    \item What is a fully general microscopic definition of anomalies? Currently, such definitions for anomalies of internal and translation symmetries have been discussed \cite{ElsePRB2014, Kapustin2024, Kawagoe2025, Liu2025, Kapustin:2025nju, LatticeAnomaly1, Shirley:2025yji, Czajka2025}, but the extension to include point-group symmetry is still largely unavailable. It is also important to generalize these definitions to continuum systems, such as particles moving in a 2D continuum with a perpendicular magnetic field that leads to Landau levels.

    \item Can anomalies be used to classify different symmetry actions? Namely, is there any criterion under which two symmetry actions are in the same class if and only if they have the same anomaly? The answer is affirmative in spin chains with unitary internal finite-group symmetries implemented by finite-depth quantum circuits \cite{Seifnashri2025, Bols2025}, but largely open for other types of symmetries and/or in other systems.

    \item 't Hooft anomalies are originally defined as obstructions to gauging symmetries. What do gauge fields of spatial symmetries mean? Ref. \cite{ThorngrenPRX2018} laid down the foundation of this subject at the level of certain coarse-grained effective theories. Can such gauge fields be formulated more microscopically, based on which one can study the renormalization group flow of the properties of the spatial symmetry defects? These questions are also related to the point that the anomaly matching framework discussed in Sec. \ref{sec: anomaly matching} relies on the crystalline equivalence principle, and it is desirable to develop a first-principles framework of anomaly matching.

    \item Based on the anomaly matching framework, multiple new quantum phases were predicted in Refs. \cite{Ye2021, Ye2023}. In which setups can these phases emerge?

\end{itemize}

% Disclosure
\section*{DISCLOSURE STATEMENT}
The authors are not aware of any affiliations, memberships, funding, or financial holdings that
might be perceived as affecting the objectivity of this review.

% Acknowledgements
\section*{ACKNOWLEDGMENTS}

LZ thanks Meng Guo, Yin-Chen He, Chong Wang, Jinmin Yi, Shiyu Zhou, and especially Ruizhi Liu and Weicheng Ye for collaborations on topics related to this review. MC thanks Maissam Barkeshli, Yang Qi and Nathan Seiberg for collaborations on related topics. LZ is supported by the National University of Singapore start-up grants A-0009991-00-00 and A-0009991-01-00. MC is supported by NSF grant DMR-2424315. This manuscript benefited from language improvements
assisted by ChatGPT 5.

% References
\bibliographystyle{ar-style4}
\bibliography{refs}

\begin{thebibliography}{107}
\expandafter\ifx\csname natexlab\endcsname\relax\def\natexlab#1{#1}\fi

\bibitem{LSM}
Lieb E, Schultz T, Mattis D. 1961.
\textit{Annals of Physics} 16(3):407--466

\bibitem{OshikawaLSM}
Oshikawa M. 2000.
\textit{Phys. Rev. Lett.} 84:1535

\bibitem{HastingsLSM}
Hastings MB. 2004.
\textit{Phys. Rev. B} 69:104431

\bibitem{PoPRL2017}
Po HC, Watanabe H, Jian CM, Zaletel MP. 2017.
\textit{Phys. Rev. Lett.} 119(12):127202

\bibitem{Watanabe2018}
{Watanabe} H. 2018.
\textit{\prb} 97(16):165117

\bibitem{Cheng2018}
{Cheng} M. 2019.
\textit{\prb} 99(7):075143

\bibitem{Ogata2018}
{Ogata} Y, {Tasaki} H. 2019.
\textit{Communications in Mathematical Physics} 372(3):951--962

\bibitem{Else_LSM_2019}
Else DV, Thorngren R. 2020.
\textit{Phys. Rev. B} 101(22):224437

\bibitem{Yao:2020xcm}
Yao Y, Oshikawa M. 2021.
\textit{Phys. Rev. Lett.} 126(21):217201

\bibitem{Ogata2020}
{Ogata} Y, {Tachikawa} Y, {Tasaki} H. 2021.
\textit{Communications in Mathematical Physics} 385(1):79--99

\bibitem{Ma2022}
{Ma} R, {Wang} C. 2023.
\textit{Physical Review X} 13(3):031016

\bibitem{Kapustin2024}
{Kapustin} A, {Sopenko} N. 2025.
\textit{Communications in Mathematical Physics} 406(10):238

\bibitem{Liu2024a}
{Liu} R, {Yi} J, {Zhou} S, {Zou} L. 2025.
\textit{\prb} 112(21):214408

\bibitem{Ma2024}
{Ma} R. 2024.
\textit{\prb} 110(10):104412

\bibitem{Liu2025}
{Liu} R, {Yi} J, {Zou} L. 2025.
\textit{arXiv e-prints} :arXiv:2510.06555

\bibitem{LuRanOshikawa2020}
Lu YM, Ran Y, Oshikawa M. 2020.
\textit{Annals of Physics} 413:168060

\bibitem{Lu2024}
Lu YM. 2024.
\textit{Annals of Physics} 470:169806

\bibitem{WuHoLu2017}
{Wu} J, {Ho} TL, {Lu} YM. 2017.
\textit{arXiv e-prints} :arXiv:1703.04776

\bibitem{XuSPTLSM}
Yang X, Jiang S, Vishwanath A, Ran Y. 2018.
\textit{Phys. Rev. B} 98(12):125120

\bibitem{JiangSPTLSM}
Jiang S, Cheng M, Qi Y, Lu YM. 2021.
\textit{SciPost Phys.} 11:024

\bibitem{Kobayashi2018}
{Kobayashi} R, {Shiozaki} K, {Kikuchi} Y, {Ryu} S. 2019.
\textit{\prb} 99(1):014402

\bibitem{Pace2024}
{Pace} SD, {Lam} HT, {Aksoy} {\"O}M. 2025.
\textit{SciPost Physics} 18(1):028

\bibitem{Pace2025}
Pace SD, Ömer M.~Aksoy, Lam HT. 2026.
\textit{SciPost Phys.} 20:007

\bibitem{Pace2026}
{Pace} SD, {Bulmash} D. 2026.
\textit{arXiv e-prints} :arXiv:2602.11266

\bibitem{Anakru2026}
{Anakru} A, {Srinivasan} S, {Li} L, {Bi} Z. 2026.
\textit{arXiv e-prints} :arXiv:2603.19189

\bibitem{Ning2026}
{Ning} SQ, {Ebisu} H, {Lam} HT. 2026.
\textit{arXiv e-prints} :arXiv:2603.19381

\bibitem{ChengPRX2016}
Cheng M, Zaletel M, Barkeshli M, Vishwanath A, Bonderson P. 2016.
\textit{Phys. Rev. X} 6(4):041068

\bibitem{JianPRB2018}
Jian CM, Bi Z, Xu C. 2018.
\textit{Phys. Rev. B} 97(5):054412

\bibitem{ChoPRB2017}
Cho GY, Hsieh CT, Ryu S. 2017.
\textit{Phys. Rev. B} 96(19):195105

\bibitem{MetlitskiPRB2018}
Metlitski MA, Thorngren R. 2018.
\textit{Phys. Rev. B} 98(8):085140

\bibitem{Ye2021}
{Ye} W, {Guo} M, {He} YC, {Wang} C, {Zou} L. 2022.
\textit{SciPost Physics} 13(3):066

\bibitem{Cheng2022}
{Cheng} M, {Seiberg} N. 2023.
\textit{SciPost Physics} 15(2):051

\bibitem{Seifnashri2023}
{Seifnashri} S. 2024.
\textit{SciPost Physics} 16(4):098

\bibitem{Liu2024}
{Liu} C, {Ye} W. 2025.
\textit{SciPost Physics} 18(5):161

\bibitem{tHooft:1979rat}
't~Hooft G. 1980.
\textit{NATO Sci. Ser. B} 59:135--157

\bibitem{Adler1969}
Adler SL. 1969.
\textit{Phys. Rev.} 177(5):2426--2438

\bibitem{Bell1969}
Bell JS, Jackiw R. 1969.
\textit{Il Nuovo Cimento A (1965-1970)} 60(1):47--61

\bibitem{Chen:2011pg}
Chen X, Gu ZC, Liu ZX, Wen XG. 2013.
\textit{Phys. Rev. B} 87(15):155114

\bibitem{CZX}
Chen X, Liu ZX, Wen XG. 2011.
\textit{Phys. Rev. B} 84(23):235141

\bibitem{ElsePRB2014}
Else DV, Nayak C. 2014.
\textit{Phys. Rev. B} 90(23):235137

\bibitem{Kawagoe2025}
{Kawagoe} K, {Shirley} W. 2025.
\textit{arXiv e-prints} :arXiv:2507.07430

\bibitem{Kapustin:2025nju}
{Kapustin} A, {Xu} S. 2025.
\textit{arXiv e-prints} :arXiv:2505.04719

\bibitem{LatticeAnomaly1}
Tu YT, Long DM, Else DV. 2026.
\textit{Phys. Rev. X} 16(1):011027

\bibitem{Shirley:2025yji}
{Shirley} W, {Zhang} C, {Ji} W, {Levin} M. 2025.
\textit{arXiv e-prints} :arXiv:2507.21267

\bibitem{Czajka2025}
{Czajka} AM, {Geiko} R, {Thorngren} R. 2025.
\textit{arXiv e-prints} :arXiv:2512.02105

\bibitem{auerbach1994interacting}
Auerbach A. 1994.
Interacting electrons and quantum magnetism.
Graduate Texts in Contemporary Physics. New York, NY, USA: Springer New York

\bibitem{WatanabePRB2018}
Watanabe H. 2018.
\textit{Phys. Rev. B} 98(15):155137

\bibitem{Aizenman2001BoundedFluctuations}
{Aizenman} M, {Goldstein} S, {Lebowitz} JL. 2001.
\textit{Journal of Statistical Physics} 103(3-4):601--618

\bibitem{Kimchi2017}
{Kimchi} I, {Nahum} A, {Senthil} T. 2018.
\textit{Physical Review X} 8(3):031028

\bibitem{Hastings2005}
{Hastings} MB, {Koma} T. 2006.
\textit{Communications in Mathematical Physics} 265(3):781--804

\bibitem{Hastings2007}
{Hastings} MB. 2007.
\textit{Journal of Statistical Mechanics: Theory and Experiment} 2007(8):08024

\bibitem{Kuwahara2019}
{Kuwahara} T, {Saito} K. 2020.
\textit{Nature Communications} 11:4478

\bibitem{Sachdev_2011}
Sachdev S. 2011.
Quantum phase transitions.
Cambridge University Press, 2nd ed.

\bibitem{Yao2018}
Yao Y, Hsieh CT, Oshikawa M. 2019.
\textit{Phys. Rev. Lett.} 123(18):180201

\bibitem{ChengPRR2020}
Cheng M, Williamson DJ. 2020.
\textit{Phys. Rev. Research} 2(4):043044

\bibitem{FeaturelessHoneycomb}
Kim P, Lee H, Jiang S, Ware B, Jian CM, et~al. 2016.
\textit{Phys. Rev. B} 94(6):064432

\bibitem{Chubukov1993}
{Chubukov} AV, {Senthil} T, {Sachdev} S. 1994.
\textit{\prl} 72(13):2089--2092

\bibitem{Chubukov1994}
{Chubukov} AV, {Sachdev} S, {Senthil} T. 1994.
\textit{Nuclear Physics B} 426(3):601--643

\bibitem{ThorngrenPRX2018}
Thorngren R, Else DV. 2018.
\textit{Phys. Rev. X} 8(1):011040

\bibitem{Gaiotto:2014kfa}
Gaiotto D, Kapustin A, Seiberg N, Willett B. 2015.
\textit{JHEP} 02:172

\bibitem{Barkeshli2014}
Barkeshli M, Bonderson P, Cheng M, Wang Z. 2019.
\textit{Phys. Rev. B} 100(11):115147

\bibitem{Affleck1988}
Affleck I, Marston JB. 1988.
\textit{Phys. Rev. B} 37(7):3774--3777

\bibitem{Wen1995}
{Wen} XG, {Lee} PA. 1996.
\textit{\prl} 76(3):503--506

\bibitem{Ran2006}
{Ran} Y, {Hermele} M, {Lee} PA, {Wen} XG. 2007.
\textit{\prl} 98(11):117205

\bibitem{Iqbal2016}
{Iqbal} Y, {Hu} WJ, {Thomale} R, {Poilblanc} D, {Becca} F. 2016.
\textit{\prb} 93(14):144411

\bibitem{He2016}
{He} YC, {Zaletel} MP, {Oshikawa} M, {Pollmann} F. 2017.
\textit{Physical Review X} 7(3):031020

\bibitem{Ranjith2019}
{Ranjith} KM, {Dmytriieva} D, {Khim} S, {Sichelschmidt} J, {Luther} S, et~al.
  2019.
\textit{\prb} 99(18):180401

\bibitem{Ding2019}
{Ding} L, {Manuel} P, {Bachus} S, {Gru{\ss}ler} F, {Gegenwart} P, et~al. 2019.
\textit{\prb} 100(14):144432

\bibitem{Bordelon2019}
{Bordelon} MM, {Kenney} E, {Liu} C, {Hogan} T, {Posthuma} L, et~al. 2019.
\textit{Nature Physics} 15(10):1058--1064

\bibitem{Hu2019}
{Hu} S, {Zhu} W, {Eggert} S, {He} YC. 2019.
\textit{\prl} 123(20):207203

\bibitem{Wietek2023}
{Wietek} A, {Capponi} S, {L{\"a}uchli} AM. 2024.
\textit{Physical Review X} 14(2):021010

\bibitem{Bag2023}
{Bag} R, {Xu} S, {Sherman} NE, {Yadav} L, {Kolesnikov} AI, et~al. 2024.
\textit{\prl} 133(26):266703

\bibitem{Karthik2015}
{Karthik} N, {Narayanan} R. 2016{\natexlab{a}}.
\textit{\prd} 93(4):045020

\bibitem{Karthik2016}
{Karthik} N, {Narayanan} R. 2016{\natexlab{b}}.
\textit{\prd} 94(6):065026

\bibitem{He2021}
{He} YC, {Rong} J, {Su} N. 2022.
\textit{SciPost Physics} 13(2):014

\bibitem{Zou2021}
{Zou} L, {He} YC, {Wang} C. 2021.
\textit{Physical Review X} 11(3):031043

\bibitem{Zeng2015}
{Zeng} B, {Chen} X, {Zhou} DL, {Wen} XG. 2015.
\textit{arXiv e-prints} :arXiv:1508.02595

\bibitem{Kitaev1997}
{Kitaev} AY. 2003.
\textit{Annals of Physics} 303(1):2--30

\bibitem{EssinPRB2013}
Essin AM, Hermele M. 2013.
\textit{Phys. Rev. B} 87(10):104406

\bibitem{Satzinger2021}
{Satzinger} KJ, {Liu} YJ, {Smith} A, {Knapp} C, {Newman} M, et~al. 2021.
\textit{Science} 374(6572):1237--1241

\bibitem{Lukin2021}
{Semeghini} G, {Levine} H, {Keesling} A, {Ebadi} S, {Wang} TT, et~al. 2021.
\textit{Science} 374(6572):1242--1247

\bibitem{Iqbal2023}
{Iqbal} M, {Tantivasadakarn} N, {Gatterman} TM, {Gerber} JA, {Gilmore} K,
  et~al. 2024.
\textit{Communications Physics} 7(1):205

\bibitem{Foss-Feig2023}
{Foss-Feig} M, {Tikku} A, {Lu} TC, {Mayer} K, {Iqbal} M, et~al. 2023.
\textit{arXiv e-prints} :arXiv:2302.03029

\bibitem{Qi2015}
{Qi} Y, {Cheng} M, {Fang} C. 2015.
\textit{arXiv e-prints} :arXiv:1509.02927

\bibitem{Qi2016}
Qi Y, Cheng M. 2018.
\textit{Phys. Rev. B} 97(11):115138

\bibitem{Vishwanath2013}
{Vishwanath} A, {Senthil} T. 2013.
\textit{Physical Review X} 3(1):011016

\bibitem{Wang2013}
{Wang} C, {Senthil} T. 2013.
\textit{\prb} 87(23):235122

\bibitem{Zou2016}
{Zou} L, {Wang} C, {Senthil} T. 2018.
\textit{\prb} 97(19):195126

\bibitem{NingPRR2020}
Ning SQ, Zou L, Cheng M. 2020.
\textit{Phys. Rev. Research} 2(4):043043

\bibitem{Ye2023}
{Ye} W, {Zou} L. 2024.
\textit{Physical Review X} 14(2):021053

\bibitem{Ye2022}
{Ye} W, {Zou} L. 2023.
\textit{SciPost Physics} 15(1):004

\bibitem{Else2020}
{Else} DV, {Thorngren} R, {Senthil} T. 2021.
\textit{Physical Review X} 11(2):021005

\bibitem{Delacretz2019}
{Delacr{\'e}taz} LV, {Hofman} DM, {Mathys} G. 2020.
\textit{SciPost Physics} 8(3):047

\bibitem{Else2025}
Else DV. 2026.
\textit{Annual Review of Condensed Matter Physics} 17(Volume 17, 2026):71--90

\bibitem{Luttinger1960}
Luttinger JM. 1960.
\textit{Phys. Rev.} 119(4):1153--1163

\bibitem{Oshikawa2000}
{Oshikawa} M. 2000.
\textit{\prl} 84(15):3370--3373

\bibitem{Senthil2002}
{Senthil} T, {Sachdev} S, {Vojta} M. 2003.
\textit{\prl} 90(21):216403

\bibitem{Senthil2003}
{Senthil} T, {Vojta} M, {Sachdev} S. 2004.
\textit{\prb} 69(3):035111

\bibitem{Bonderson2016}
{Bonderson} P, {Cheng} M, {Patel} K, {Plamadeala} E. 2016.
\textit{arXiv e-prints} :arXiv:1601.07902

\bibitem{Bonetti2026}
Bonetti PM, Christos M, Nikolaenko A, Patel AA, Sachdev S. 2026.
\textit{Reports on Progress in Physics} 89(4):044501

\bibitem{Ma2023}
{Ma} R, {Zhang} JH, {Bi} Z, {Cheng} M, {Wang} C. 2025.
\textit{Physical Review X} 15(2):021062

\bibitem{Panahi2026}
{Panahi} Y, {Sahu} S, {Manjunath} N, {Wang} C. 2026.
\textit{arXiv e-prints} :arXiv:2602.02648

\bibitem{HsiehPRL2016}
Hsieh TH, Hal\'asz GB, Grover T. 2016.
\textit{Phys. Rev. Lett.} 117(16):166802

\bibitem{OmerPRB2021}
Aksoy OM, Tiwari A, Mudry C. 2021.
\textit{Phys. Rev. B} 104(7):075146

\bibitem{HuangPRB2017}
Huang SJ, Song H, Huang YP, Hermele M. 2017.
\textit{Phys. Rev. B} 96(20):205106

\bibitem{Seifnashri2025}
{Seifnashri} S, {Shirley} W. 2025.
\textit{arXiv e-prints} :arXiv:2503.09717

\bibitem{Bols2025}
{Bols} A, {De Roeck} W, {De Wilde} M, {Carvalho} BdO. 2025.
\textit{Communications in Mathematical Physics} 407(1):10

\end{thebibliography}

\end{document}